\documentclass[final,5p,times,authoryear, twocolumn]{elsarticle}
\usepackage{packages}

\DeclareUnicodeCharacter{0301}{\'{i}}
\DeclareUnicodeCharacter{0456}{\textit{i}}
\DeclareUnicodeCharacter{FFFD}{?}
\DeclareUnicodeCharacter{2082}{\textsubscript{2}}

\makeatletter
\def\ps@pprintTitle{%
	\let\@oddhead\@empty
	\let\@evenhead\@empty
	\let\@oddfoot\@empty
	\let\@evenfoot\@oddfoot
}

\title{Digital Twin for Autonomous Surface Vessels: Enabler for Safe Maritime Navigation}

\begin{document}
\begin{frontmatter}

\author{Daniel~Menges\corref{cor1}\fnref{address1}}
\ead{daniel.menges@ntnu.no}
\cortext[cor1]{Corresponding author}
\author[address1,address2]{Adil Rasheed}
\ead{adil.rasheed@ntnu.no}

\address[address1]{Department of Engineering Cybernetics, Norwegian University of Science and Technology, Trondheim, Norway}
\address[address2]{Mathematics and Cybernetics, SINTEF Digital, Trondheim, Norway}

\begin{abstract}
Autonomous surface vessels (ASVs) are becoming increasingly significant in enhancing the safety and sustainability of maritime operations. To ensure the reliability of modern control algorithms utilized in these vessels, digital twins (DTs) provide a robust framework for conducting safe and effective simulations within a virtual environment. Digital twins are generally classified on a scale from 0 to 5, with each level representing a progression in complexity and functionality: Level 0 (Standalone) employs offline modeling techniques; Level 1 (Descriptive) integrates sensors and online modeling to enhance situational awareness; Level 2 (Diagnostic) focuses on condition monitoring and cybersecurity; Level 3 (Predictive) incorporates predictive analytics; Level 4 (Prescriptive) embeds decision-support systems; and Level 5 (Autonomous) enables advanced functionalities such as collision avoidance and path following.
These digital representations not only provide insights into the vessel's current state and operational efficiency but also predict future scenarios and assess life endurance. By continuously updating with real-time sensor data, the digital twin effectively corrects modeling errors and enhances decision-making processes. Since DTs are key enablers for complex autonomous systems, this paper introduces a comprehensive methodology for establishing a digital twin framework specifically tailored for ASVs. Through a detailed literature survey, we explore existing state-of-the-art enablers across the defined levels, offering valuable recommendations for future research and development in this rapidly evolving field.
\end{abstract}

\end{frontmatter}
	
\section{Introduction} \label{sec:introduction}
Commercial shipping is a cornerstone of the global economy, enabling the transport of vast quantities of goods across oceans. In 2020, the maritime sector moved approximately 10,648 million tonnes of seaborne goods, emphasizing its critical role in international trade \citep{2021a}. However, the industry's rapid growth has also contributed to significant environmental challenges. Shipping currently accounts for over 1 billion tonnes of carbon dioxide emissions annually, representing roughly 2.5\% of global emissions \citep{IMO2020}. Due to rising environmental concerns, developing cost-effective and environmentally sustainable solutions in commercial shipping has become an urgent priority for many countries.

One promising solution is autonomous operation. Autonomous vessels have the potential to revolutionize maritime operations by increasing efficiency and reducing environmental impact. These vessels can follow optimal navigational routes, maximize fuel efficiency, and lower CO2 emissions. Additionally, since human error is responsible for nearly 80\% of maritime accidents \citep{SanchezBeaskoetxea2021}, autonomy could significantly enhance safety by satisfying strict safety protocols, detecting hazards beyond human capability, and making real-time decisions without fatigue. 

Despite these potential benefits, the transition to autonomous shipping introduces a range of technological challenges. Implementing autonomous surface vessels (ASVs) in real-world maritime environments, particularly in high-stakes scenarios such as deep-sea shipping with large cargo, poses substantial risks. Minor technological errors in such settings can lead to devastating accidents, environmental disasters, and financial losses. To mitigate these risks, the concept of a digital twin has emerged as a game-changing technology.

A digital twin generates a virtual model of a vessel and its operating environment, enabling the simulation and evaluation of operational strategies prior to real-world implementation. This capability significantly reduces the risks associated with the introduction of autonomous systems by offering a safe testing ground for innovative technologies. By simulating a wide range of scenarios, digital twins can accelerate the deployment of autonomous shipping operations that are not only safer and more sustainable but also more cost-effective. However, viewing digital twins merely as computer-aided design (CAD) models or basic simulators is a narrow and restrictive perspective. To fully exploit the potential of digital twins in autonomous shipping, a more refined and context-specific understanding of the concept is essential. 
\newpage
The current work seeks to address this by the following:

\begin{itemize} 
    \item \textit{Defining digital twin and its capability levels for maritime applications:} Digital twins are categorized into six levels (0 to 5), each representing increasing levels of sophistication. 
    \item \textit{Overview of current relevant technologies:} A comprehensive overview of the current state of technologies essential for creating digital twins for ASVs is provided. These include advances in sensor integration, condition monitoring, predictive analytics, and advanced collision avoidance systems. 
    \item \textit{Identification of opportunities and challenges:} The potential benefits and challenges of developing highly functional digital twins for autonomous shipping are identified. 
    \item \textit{Recommendations for stakeholders:} Recommendations are offered to various stakeholders, outlining how they can maximize the value derived from digital twin technology in autonomous maritime systems in return of their contributions. These suggestions highlight priority areas for future research and practical strategies for technology development. 
\end{itemize}

For a better organization of the review and related perspective, this article is structured into five sections. The current introduction section outlined the significance of autonomous shipping and the necessity of digital twin technology to address current challenges. The following Section~\ref{sec:DigitalTwins} explains the general concept of a digital twin, detailing its progression from Level 0 (standalone systems) to Level 5 (fully autonomous operations). In Section~\ref{sec:Methodlogy_for_literature_survey}, the approach taken to compile and analyze existing research is described, ensuring a thorough examination of relevant studies. Section~\ref{sec:existing_state_of_the_art_enablers} is divided into subsections that explore each level of digital twin capability, from Level 0-Standalone to Level 5-Autonomous, discussing the advancements and applications at each stage. Finally, the article concludes with Section~\ref{sec:conclusionsandrecommendations}, summarizing key findings and proposing future directions for research in digital twin technology for autonomous maritime operations.

\section{Digital Twins} \label{sec:DigitalTwins}
\begin{figure*}[ht]
	\centering
	\includegraphics[width=\linewidth]{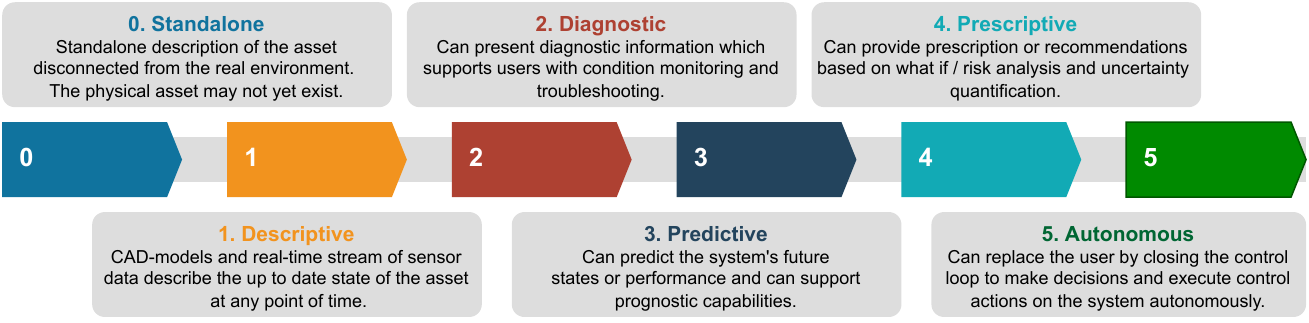}
	\caption{Capability levels of digital twins.}
	\label{fig:DT_capability_levels}
\end{figure*}

Numerous definitions of digital twins exist within the literature, ranging from simple CAD models used primarily for visualization \citep{Pal2022} to high-fidelity simulators for design optimization and planning. Initially, the limitations of expensive or poor-quality sensors, lack of real-time data, and the absence of computationally efficient simulators confined the use of digital twins largely to the design phase. However, technological advancements are transforming the scope and application of digital twins. Improvements in cost-effective sensors, enhanced communication technologies, the ability to collect high-quality data, and  a boost in computational efficiency and accuracy of models driven by artificial intelligence are extending the relevance of digital twins throughout the full lifecycle of assets and even beyond. This is enabling dynamic real-time interaction between the physical asset and its digital counterpart through data exchange and communication \citep{Pal2022, Grieves, Wu2020, Stadtmann2023doa, Stadtmann2023dif}. Moreover, more recent definitions of digital twins attempt to extend the concept to include prediction, optimization, and enhanced decision-making capabilities \citep{Rasheed2020dtv, Reiche2021, Wunderlich2021}. The current article will be based on the following definition of digital twins provided in \citep{Rasheed2020dtv}.
\newline
\newline
\fbox{
\parbox{0.95\linewidth}{
A {\bf digital twin} is defined as a virtual representation of a physical asset or a process enabled through data and simulators for real-time prediction, optimization, monitoring, control, and informed decision-making.}
}
\vspace{3pt}
\newline
The definition was endorsed in the context of wind energy by major industry players active in digital twin-related research \citep{Stadtmann2023dti} and has shown to be working well in other application areas like built environment \citep{Elfarri2023aida} and aquaculture \citep{Fore2024dti}. We build upon the same definition and adapt the capability levels of digital twins, presented in Fig.~\ref{fig:DT_capability_levels}, initially proposed in \cite{DNV-RP-A204} and later refined by \cite{Stadtmann2023dti}. The capability level scales are elaborated in Section~\ref{subsec:level0} to Section~\ref{subsec:level5} and will form the basis for structuring the following literature survey.

    \subsection{Capability Level 0: Standalone}
    \label{subsec:level0}
    A digital twin at capability level 0 is referred to as a standalone digital twin, primarily characterized by the use of three-dimensional (3D) geometric representations of an asset and simulations to model its behavior. At this stage, the digital twin may not have an established data connection to the actual asset, and in some cases, the asset may not even exist yet. While a standalone digital twin can be sufficient for assessing the design and operational characteristics of an asset, it is inadequate for applications requiring real-time data integration, such as autonomous systems. However, developing a standalone digital twin offers significant advantages during the initial design phases. It allows for early evaluation of design, production, and operational processes, helping to identify potential engineering issues and guiding subsequent design iterations. This foundational work sets the stage for more advanced digital twin capabilities, which incorporate dynamic interactions with the physical world, ultimately improving the asset's overall performance and safety.  
    
    \subsection{Capability Level 1: Descriptive}
    \label{subsec:level1}
    At capability level 1, a descriptive digital twin establishes a unidirectional data stream from the physical asset to its digital counterpart, enabling real-time monitoring and representation of the asset’s behavior. A range of sensors track both internal performance metrics and external environmental conditions. However, the data is often challenged by the "5Vs" of big data: volume, velocity, veracity, variety, and validation. Additionally, certain quantities of interest may not be directly measurable and must be inferred from more easily measured quantities.
    
    To address these challenges, numerical models are employed to refine and fuse the data, enhancing its resolution, reducing its volume, and improving the reliability and accuracy of the insights. These models can also help manage the high velocity of incoming data by filtering and prioritizing information, while dealing with variety and veracity by integrating diverse data sources and compensating for uncertainties. Through this process, models bridge gaps in the data, enabling the estimation of key performance indicators that are otherwise difficult to measure directly.
    
    This ability to handle the 5Vs and generate actionable insights from real-time data streams is what defines a descriptive digital twin. By transforming raw data into valuable information, the descriptive digital twin lays the groundwork for higher-level digital twin capabilities that can further optimize asset performance and decision-making.

    \subsection{Capability Level 2: Diagnostic}
    \label{subsec:level2}
    At capability level 2, diagnostic capabilities are essential for real-time monitoring and evaluation of an asset's condition. A diagnostic digital twin utilizes models and sensor data to generate actionable insights. At the core of this level is condition monitoring, which tracks the real-time health and performance of the asset and its components. This real-time assessment allows for immediate identification of issues and inefficiencies, ensuring that the asset operates within a safe range. In addition to condition monitoring, cybersecurity plays a crucial role in safeguarding the integrity of both the digital twin and the asset’s systems. As assets rely on complex, interconnected networks of sensors and control algorithms, the potential for cyber threats increases. A diagnostic digital twin continuously monitors the cyber environment, detecting vulnerabilities, identifying anomalies, and taking preventive measures to protect against data breaches, system manipulation, and malicious intrusions. By integrating these cybersecurity diagnostics, the digital twin maintains operational safety and the confidentiality of sensitive data.
    
    \subsection{Capability Level 3: Predictive}
    \label{subsec:level3}
    A predictive digital twin focuses on anticipating future states of an asset since no data exists beyond the current time. The core of this capability lies in developing advanced forecasting models that can predict an asset's future behavior and condition. These predictions enable critical applications such as predictive maintenance, where potential failures or degradations are identified before they occur, and process optimization, allowing systems to operate more efficiently by preemptively addressing issues or optimizing performance based on future conditions.
    
    For predictive digital twins to be effective, several key characteristics are essential in the models used. Accuracy is paramount to ensure reliable predictions, while minimizing uncertainty helps reduce risk in decision-making. Computational efficiency is crucial to provide timely forecasts without excessive resource consumption. Transparency in model behavior and assumptions enhances trust and interpretability, especially when dealing with complex systems. Finally, generalizability ensures that models can be applied to a wide range of conditions or assets, making them versatile and adaptable to different scenarios. By balancing these factors, predictive digital twins become powerful tools for improving asset management and operational efficiency.
    
    \subsection{Capability Level 4: Prescriptive}
    \label{subsec:level4}
    A prescriptive digital twin is defined by its ability to simulate various "what-if" scenarios. This capability builds upon the foundation of earlier digital twin levels. In complex systems, prescriptive capabilities enable the simulation of multiple operational strategies to identify options that minimize risk and enhance efficiency. For example, it can optimize processes based on factors such as environmental conditions, system interactions, and external influences, using advanced modeling techniques.
    
    To achieve this, the digital twin incorporates real-time and historical data, along with external factors like environmental forecasts or data from interconnected systems. It can then provide optimal recommendations, such as adjusting operational strategies to avoid potential hazards or inefficiencies. This capability holds significant potential for improving decision-making and performance across various sectors.
    
    \subsection{Capability Level 5: Autonomous}
    \label{subsec:level5}
    At capability level 5, a fully autonomous digital twin establishes bidirectional communication with the asset, creating a closed control loop where the asset continuously updates the digital twin with real-time data, and in return, the digital twin autonomously controls the asset. For this system to function effectively, the foundational capabilities developed in levels 1 through 4 must be robust. Any weaknesses at earlier stages could result in critical misinterpretations, potentially causing harm to the physical asset. An autonomous digital twin not only responds to real-time conditions but also anticipates future scenarios, enabling it to make informed control decisions for optimal outcomes. The ability to forecast and act on future scenarios is key to enhancing performance and ensuring smooth operations.
    
    However, safety must remain the top priority. The digital twin must maintain an accurate understanding of the asset's internal states and external environment to avoid dangerous situations. If the digital twin's situational awareness is limited, it could transmit faulty control commands, leading to operational failures or safety risks. 

\section{Methodology for Conducting the Literature Survey} \label{sec:Methodlogy_for_literature_survey}
Building on the understanding of digital twins and their capability levels as discussed in the previous section, we now explore the current state of research on digital twins in the context of autonomous surface vessels (ASVs). Conducting a thorough literature review is a crucial step in assessing the latest advancements in this field. This approach involves systematically gathering, evaluating, and synthesizing relevant studies across the interdisciplinary areas of digital twin technology, maritime automation, and autonomous navigation systems. Through this process, we aim to gain a comprehensive view of the current state of the art in ASV-related digital twin research.

The literature survey begins with a comprehensive search strategy designed to capture a wide range of relevant studies. The search strategy involved several key steps:

\begin{enumerate} 
\item \textit{Defining Search Queries:} A set of search queries is developed based on the research questions or hypotheses. These queries should include relevant keywords, phrases, and Boolean operators. For example, the search command could be structured as follows: \textit{("keyword~1" AND "keyword~2") OR ("keyword~3" AND "keyword~4")}. To cover a wide range of research on "Autonomous Surface Vessels," the keyword constellation \textit{("vessel" OR "ship" OR "boat" OR "surface vehicle") AND ("autonomous" OR "unmanned")} ensures inclusivity by capturing various terms used for watercraft (vessel, ship, boat, surface vehicle) and their autonomy (autonomous, unmanned), thereby encompassing different terminologies across diverse studies. 
The core keywords (see blue box) were primarily combined and coupled with the alternative keywords (green box) to form comprehensive search prompts. Additionally, the core keywords were cross-referenced with specific alternative keyword groups, enhancing both the breadth and focus of the search strategy.

\begin{tcolorbox}[colback=blue!5!white, colframe=blue!50!black, title=Core Keyword Prompt]
\begin{itemize}
  \item "Digital Twin" 
  \item ("Vessel" OR "Ship" OR "Boat" OR "Surface Vehicle") AND ("Autonomous" OR "Unmanned")
\end{itemize}
\end{tcolorbox}

\begin{tcolorbox}[colback=blue!5!white, colframe=green!30!black, title=Alternative Keywords (AND)]
\begin{itemize}
    \item "Control"
    \item "Big Data"
    \item "Cybersecurity"
    \item "Path Following"
    \item "Collision Avoidance"
    \item "Sensors"
    \item "Sensor Fusion"
    \item "Situational Awareness"
    \item "Target Tracking"
    \item "Condition Monitoring"
    \item "State Estimation"
    \item "COLREG"
    \item "Maritime Standards"
    \item ("Sea State" OR "Environment")
    \item ("Prediction" OR "Forecast")
    \item (("Data-Driven" OR "Physics-Based" OR "Hybrid" OR "Reduced-Order" OR "3D") AND ("Modeling" OR "Modelling"))
\end{itemize}
\end{tcolorbox}

\item \textit{Selecting Databases:} In order to ensure comprehensive coverage of relevant literature, we selected and utilized widely recognized academic databases such as Scopus, Web of Science, ScienceDirect, and IEEE Xplore. These platforms offer a diverse range of interdisciplinary research, with Scopus and Web of Science being particularly known for their citation tracking and multidisciplinary coverage. For a detailed comparison between Scopus and Web of Science, refer to the work by \cite{singh_journal_2021}, which highlights their strengths in citation metrics and journal coverage. A concise comparison of the selected databases is presented in Table~\ref{tab:database_comparison}.

\begin{table*}[t!]
\centering
\caption{Comparison of Scopus, Web of Science, ScienceDirect, and IEEE Xplore.}
\label{tab:database_comparison}
\resizebox{\textwidth}{!}{%
\renewcommand{\arraystretch}{1.5} 
\begin{tabular}{|p{3cm}|p{2.5cm}|p{5.5cm}|p{3.5cm}|p{3cm}|}
\hline
\textbf{Database} & \textbf{Type} & \textbf{Focus} & \textbf{Strengths} & \textbf{Weaknesses} \\ \hline \hline
\textbf{Scopus} & Citation tracking & Multidisciplinary & Broad coverage & No full-text \\ \hline
\textbf{Web of Science} & Citation tracking & Multidisciplinary & Focus on high-impact journals & No full-text \\ \hline
\textbf{ScienceDirect} & Full-text & Engineering, Natural Sciences, Health Sciences, Social Sciences and Humanities & Full-text of Elsevier journals & Limited to Elsevier publications \\ \hline
\textbf{IEEE Xplore} & Full-text & Engineering, Computer Sciences, Natural Sciences, Mathematics, Technical Communications, Education, Management, Law, Policy & Extensive collection of IEEE papers, strong in technology and engineering research & Limited to IEEE content \\ \hline
\end{tabular}%
}
\end{table*}

\item \textit{Refining Search Results:} Finally, the search results were reviewed and refined to remove duplicates and non-relevant studies. This process involved scanning titles, abstracts, and keywords to identify potentially pertinent studies for full-text review.
\end{enumerate}

\section{Existing State of the Art of Enablers for Digital Twins}
\label{sec:existing_state_of_the_art_enablers}

\subsection{Level 0: Standalone}
Standalone digital twins for autonomous ships are primarily based on CAD models and high-fidelity physics-based simulators. These simulators serve as representative models of the ship's systems and behavior. In this stage, data-driven models have limited application since real-time data from the asset is typically unavailable. However, surrogate models are occasionally developed to assist in design optimization, providing faster approximations while maintaining sufficient accuracy to inform early design decisions. These models support the evaluation of different design configurations without relying on actual operational data.

\subsubsection{3D Modeling}
There are various methods to create a three-dimensional (3D) model, each tailored to different application requirements. Several 3D modeling software environments use B-splines (basis splines) for compact and precise representation. B-splines are a powerful tool for constructing smooth curves and surfaces, enabling designers to efficiently define geometries with minimal control points. In CAD programs, B-splines are often employed to build object geometries in a virtual space. The outcome is a detailed visualization of an object represented through a mathematical mesh of curves and surfaces. CAD-based models are essential for generating accurate representations of solid bodies and are widely used in fields such as mechanical engineering, product design, and architecture.

\subsubsection{Simulation Models}
Modeling and simulation are fundamental to the development and implementation of a standalone digital twin. At this capability level, the majority of models are knowledge-based or physics-based, grounded in first principles and governed by physical laws. While these models may not achieve perfect accuracy due to simplifying assumptions, approximations, and incomplete representation of the full governing physics, they serve as effective representative models. In addition, optimization algorithms are used in combination with these models. The following section provides an overview of the physics-based models relevant to maritime applications.

\paragraph{Finite Element Modeling}
Finite element methods (FEMs) are used to analyze structural integrity by simulating how materials deform and behave under dynamic loads, such as plastic and elastic deformations \citep{Kitamura2002fat}. FEM discretizes complex structures, like surface vessels, into smaller elements (e.g., tetrahedrons or hexahedrons) to approximate how they respond to various forces. This is crucial for predicting structural behavior in real-world maritime conditions, where vessels are subjected to dynamic stresses from waves, wind, and operational loads. In the context of surface vessels, FEM tools integrated with CAD systems allow engineers to simulate stress, strain, and other physical factors, enhancing the accuracy of the design process and ensuring the vessel's structural integrity under demanding conditions. 
    
\paragraph{Ship Hydrodynamic Simulations} 
These simulate the behavior of a ship in water, considering factors like wave conditions and sea currents. This helps in understanding how a vessel will respond to different environmental conditions. In \cite{Lee2022rtd}, a real-time digital twin for ship operations is presented, integrating wave prediction and hydrodynamic analysis to forecast ocean conditions, ship responses, and optimal routes, enhancing operational risk management and performance in seaways. \cite{Han2020ass} explore its application in marine operations by identifying key hydrodynamic model parameters for adaptive tuning based on vessel motion sensitivity. An extended algorithm is presented by \cite{Han2021vhm} for real-time tuning of vessel hydrodynamic model parameters, utilizing onboard sensor data, spectral analysis, and Bayesian updating to improve vessel motion prediction accuracy while maintaining computational efficiency.
    
\paragraph{Ship Maneuvering Simulations}
Simulations of ship maneuvers, such as docking, undocking, and navigating through narrow channels is paramount. This aids in training and evaluating the skills of ship operators. \cite{Zyczkowski2020srp} outlines a methodology for determining multi-criteria routes for sailing ships by discretizing sea areas and vessel properties, defining user requirements, and using an algorithm to recommend routes based on specific criteria and user categories. \cite{Xu2023msp} present a multi-stage geometric path planning algorithm for automated unmanned surface vessel (USV) recovery, allowing for various initial conditions and velocities, and demonstrates through simulation that it effectively achieves recovery in both straight and curved courses. An evolutionary potential field model for trajectory planning of uncrewed ships is introduced by \cite{Zheng2024ubp}, combining it with a differential evolution algorithm and a quadratic optimization smoothing algorithm to improve path design and limit turning angles, demonstrating its effectiveness through simulation experiments.

\paragraph{Weather-Driven Path Planning} Weather forecasts play a crucial role in optimizing ship routes by considering wind, waves \citep{Lee2022rtd}, and currents, which enhances fuel efficiency, safety, and ensures timely arrivals. An optimal path planning approach for USVs is proposed by \cite{Xu2022gpp}, featuring a novel risk-penalty-related A* algorithm that integrates boat weight and marine weather. This method, validated through Unreal Engine simulations, demonstrates improved efficiency and safety by generating adaptable navigation paths under varying conditions. Similarly, \cite{Artusi2021spp} present a Deep Reinforcement Learning (DRL) model for ship path planning that incorporates weather forecasts and accounts for simplified static and mobile obstacles. Although the model shows some adaptability to changing conditions, further refinement is needed to address the complexities of dynamic environments and improve overall navigation performance.
   
\paragraph{Ship Speed Models for Fuel Optimization}
Determining the most fuel-efficient speed for a voyage involves balancing fuel costs, time constraints, and environmental regulations. High fuel consumption and emissions have driven the development of speed optimization strategies, with research showing that reducing speed significantly lowers fuel consumption and emissions, as illustrated in case studies of the MV Meratus Mamiri on Indonesian routes \citep{Gusti2016teo}. \cite{Lundh2016eao} propose an online method for estimating the specific fuel consumption of individual diesel generators, optimizing power plant operations and achieving 4-6\% fuel savings by addressing variations in fuel use across generators over time. This method was validated using data from a cruise ship. Additionally, optimizing ship speeds on routes with port time windows, by solving a shortest path problem on a directed acyclic graph, can lead to significant fuel savings and emissions reductions \citep{Fagerholt2010rfe}. In \cite{Qi2012mfe}, a vessel scheduling problem is formulated to minimize expected fuel consumption and emissions on liner shipping routes, accounting for uncertainties in port times and frequency requirements. Simulation-based stochastic approximation methods are used, with specific cases demonstrating the convexity and differentiability of the objective function, offering valuable insights into the impact of port uncertainties.

\paragraph{Collision Risk Models} 
Collision risk assessment involves modeling the movements of multiple vessels in close proximity to evaluate potential hazards. Accurate modeling of vessel trajectories and interactions is essential for the development and implementation of effective collision avoidance strategies. By predicting the relative positions and velocities of nearby vessels, these models help identify potential collision risks, allowing for timely adjustments in navigation or speed to ensure safe operations. One approach could involve utilizing Automatic Identification System (AIS) data from multiple vessels to simulate and evaluate collision risks, thereby enhancing situational awareness. Collision risk for ferries in the Yangtze River is assessed using AIS data, incorporating risk influencing factors such as distance to the closest point and time to the closest point, with weights determined by entropy theory and a proposed collision risk index enabling real-time risk assessment and identification of high-risk areas \citep{Cai2021cra}. Given the challenge of analyzing offshore supply vessel collisions with platforms in uncertain environments and the limitations of traditional fault tree analysis due to lack of failure data, \cite{John2017crm} propose using fuzzy fault tree analysis to integrate and synthesize diverse data, offering a flexible framework for evaluating collision risks and informing resource deployment decisions. To address the persistent threat of ship collisions in busy ports and waterways, \cite{Li2023rtc} present an integrated approach that combines regional gridding, a risk model based on accident data, and a real-time risk model using random forest to identify high-risk areas and enhance maritime safety management, as demonstrated in a case study at Shenzhen port. Moreover, \cite{Wang2024scr} introduce a multi-criteria decision-making framework based on Dempster–Shafer evidence theory for ship collision risk modeling, integrating AIS data and expert judgments to assess collision risks in major global waterways, with results highlighting elevated risks in channels.
    
\paragraph{Port Operations}
Simulating vessel movements within a port is crucial for optimizing traffic flow and reducing congestion, while modeling container terminal operations helps streamline handling processes, minimize waiting times, and improve overall efficiency \citep{Fagerholt2010rfe, Qi2012mfe, Li2023rtc}. Ports with dense vessel traffic pose a higher risk of collisions, and the use of AIS data allows for precise analysis of ship movements to enhance traffic management and optimize waterway design. In \cite{Mou2010soc}, a study focused on the traffic separation scheme at Rotterdam Port utilizes AIS data and linear regression models to assess collision risks, introducing a dynamic method for real-time traffic monitoring and risk management.
    
\paragraph{Emergency Response and Safety}
Effective emergency response and safety protocols are critical in maritime environments, where rapid, well-coordinated actions can significantly mitigate risks and safeguard both human lives and assets amidst unpredictable and hazardous conditions. Emergency preparedness drills at sea are essential but carry significant risks, particularly due to human error, which is often overlooked. A hybrid approach is demonstrated in \cite{Ahn2022aoa}, utilizing fuzzy decision-making methods to assess human reliability and evaluate risks during emergency drills using a rescue boat drill as a case study.
    
\paragraph{Environmental Impact Assessments}
Simulating oil spill dispersion and vessel emissions enables the assessment of environmental impacts, supporting the development of mitigation strategies and regulatory compliance measures. A coupled oil fate and effects model estimates the distribution and biological impact of spilled oil on habitats, wildlife, and aquatic organisms by simulating physical processes and exposure, with case studies like the Exxon Valdez spill validating its accuracy \citep{French-McCay2004osi}. Besides, \cite{Wu2021aqd} introduce a quantitative decision-making model for early emergency response to ship oil spills, addressing challenges such as time constraints and limited resources by integrating a hierarchical framework and evidential reasoning to select the best response action.
    
\paragraph{Training Simulators}
Docking and bridge simulators offer ship officers realistic training environments to practice navigation, communication, and emergency response, enhancing their readiness for real-world maritime operations.
The validation study by \cite{Lauronen2020vov} assesses a virtual reality ship command bridge against maritime training simulator standards, revealing that while it currently falls short of all criteria, advancements in hardware and programming could address these issues, confirming Virtual Reality (VR) as a valid and efficient tool for command bridge training. Furthermore, in assessing the effectiveness of stress evaluation methods for ship navigators during critical decision-making scenarios, salivary amylase activity has emerged as a promising index, demonstrating its utility in both simulator and real-world settings for safe navigation training \citep{Murai2009eme}.

\subsubsection{Maritime Standards} \label{sec: Standardization}
The implementation of autonomous ships necessitates regulations and standardization, which can be incorporated into a standalone digital twin. Since autonomous ships are mainly relevant for the public sector, the private sector will cover mostly human-driven ships. Considering the number of accidents caused by human faults, autonomous ships have to adapt to these mistakes. In the maritime field, different standardization approaches are proposed. The IMO is working on Maritime Autonomous Surface Ship (MASS) strategies \citep{IMO}. Generally, the international regulations for preventing collisions at sea (COLREG) provide guidelines for collision avoidance \citep{InternationalMaritimeOrganization}.

\subsection{Level 1: Descriptive}
Digital twins rely heavily on the availability of real-time data, and in the context of autonomous vessels, this data is crucial for creating situational awareness. A descriptive digital twin serves as the foundation for all higher-level digital twin capabilities. In autonomous vessels, the data is often diverse, including inputs such as acoustic signals, images, pressure, temperature, LiDAR, and RADAR. Advanced algorithms are required to fuse these diverse data sources and extract meaningful information to support decision-making processes. Additionally, certain quantities of interest may be difficult or too expensive to measure directly, necessitating their estimation from other, more easily measurable parameters. This is where modeling becomes essential. Unlike in standalone digital twins, where data-driven models had limited application, descriptive digital twins benefit from the combined use of physics-based and data-driven models. In fact, there is growing interest in hybridizing these models to enhance their effectiveness. All these aspects are discussed in the following section. 

\subsubsection{Sensor technologies}
While conventional ships already have multiple sensors, such as fuel indicators, oil temperature sensors, oil pressure sensors, engine speed sensors, etc., the number of sensors for autonomous ships enlarges. 
Autonomous ships rely heavily on a suite of sensor technologies to perceive their surroundings and navigate safely. These sensors act as the eyes and ears of the vessel, feeding data to the onboard AI systems for real-time decision-making. In the following, a breakdown of the key sensor types crucial for autonomous ships is listed.

\paragraph{Positioning and Navigation Sensors}
\begin{itemize}
    \item Global Navigation Satellite System (GNSS) receivers: GNSS has been a common acquisition source for localizing the ship's position, providing precise positioning data using signals from global positioning system (GPS) or other satellite constellations. However, global navigation satellite systems (GNSS) are vulnerable to radio frequency interference (RFI), which can disrupt signal reception and degrade positioning accuracy. This interference can arise from various sources, including intentional jamming and unintentional interference, posing significant challenges for reliable navigation and safety-critical applications \citep{Dempster2016ilf}.
    \item Automatic Identification System (AIS): AIS messages can provide information about multiple vessels in the area, with data fields such as maritime mobile service identity (MMSI), coordinated universal time (UTC) timestamp, course over ground (COG), speed over ground (SOG), and world geodetic system (WGS84) latitude and longitude. However, AIS is typically used by commercial ships, and the timestamp is not always locally monotonic for a single MMSI, which requires filtering methods for remedy \citep{Wilthil2017}. Furthermore, sometimes commercial fishing vessels turn off the public tracking system as fishers do not want to reveal their fishing location.
    \item Inertial Measurement Unit (IMU): IMUs are commonly used for detecting the translational and rotational motion of technical systems. An IMU consists of an accelerometer, gyroscope, and magnetometer, providing precise information about a ship’s acceleration and orientation \cite{DosSantos2013ano}. However, while they offer high-resolution data for short-term navigation, they are susceptible to drift over time, which can affect long-term accuracy without correction from external references. 
\end{itemize}

\paragraph{Obstacle Detection and Ranging Sensors}
\begin{itemize}
    \item Cameras: These passive sensors extract visible light from the electromagnetic spectrum, offering high spatial accuracy for object identification. Cameras can be employed for both object classification and tracking in various applications \citep{Yilmaz2006ota}. However, cameras face limitations under poor lighting and adverse weather conditions, and their computational effort is often high \citep{Thombre2022}. 
    \item Radio Detection and Ranging (RADAR): RADAR is an active sensor that uses frequency-modulated waveforms to detect moving targets. It operates effectively in all weather and lighting conditions but has a relatively wide beam, making structural detail identification difficult \citep{Thombre2022}. However, RADAR provides accurate distance measurements and velocity information via the Doppler effect and is widely applied in classification and tracking tasks \citep{Brusch2011ssw}.
    \item Light Detection and Ranging (LiDAR): LiDAR sensors emit pulsed laser beams to measure range with narrow beams and high resolution. While highly accurate, they are susceptible to weather phenomena such as precipitation, which may distort the readings \citep{Thombre2022}. However, LiDAR can capture detailed three-dimensional information about the environment, enabling precise mapping and object recognition.
    \item Sound Navigation and Ranging (SONAR): SONAR systems use sound waves to detect underwater objects and measure depth, commonly used for avoiding underwater obstacles like reefs and wrecks \citep{Hong2016aus}. Additionally, SONAR can effectively track ship wakes \citep{Lei2014ros}, providing valuable data for navigation and monitoring. While SONAR is highly effective in murky waters where optical systems fail, it may struggle with long-range detection due to sound attenuation in the water.
\end{itemize}

\paragraph{Environmental Sensors}
\begin{itemize}
    \item Weather Sensors: Anemometers measure wind speed and direction, which, when combined with wave sensors, help assess sea states \citep{Rahmstorf1989, Gavrikov2021}. Anemometer readings also indirectly contribute to forecasting wave behavior since wind influences waves. However, using weather forecasts can further improve environmental perception over both short and long horizons.
\end{itemize}

\paragraph{Additional Sensors}
\begin{itemize}
    \item Microphone Arrays: Emerging technologies such as microphone arrays are being utilized to detect and identify a range of sounds, including port noises \citep{Bocanegra2022ana} and distress signals from vessels. Additionally, these systems enhance the capability for gas leak detection \citep{Li2021hsg}. 
    \item Infrared Cameras: These cameras detect thermal radiation, which is useful for night operations and low-visibility conditions. They are effective for both detection tasks \citep{Dulski2011dos} and condition monitoring of components under thermal stress \citep{bagavathiappan_infrared_2013, Menges2024cam, Menges2024rtp}.
\end{itemize}

\begin{figure*}[t!]
	\centering
	\includegraphics[width=\textwidth]{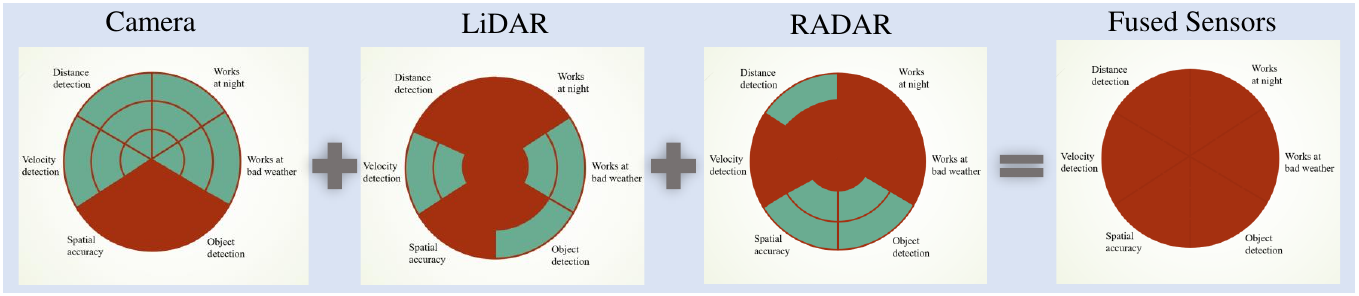}
	\caption{Principle of sensor fusion demonstrated using the most widely used perception sensors in robotics.}
	\label{fig:Sensors_fusion}
\end{figure*}

\paragraph{Sensor Fusion}
It is important to note that these sensors work best when combined through sensor fusion. By merging data from various sensors, autonomous ships can get a more comprehensive picture of their environment, leading to safer and more efficient navigation. Cameras have high spatial accuracy. Thus, they are capable of identifying objects accurately. However, cameras are unable to track objects at night and in challenging weather conditions. In comparison, RADARs can operate during all weather conditions and during every time of the day. While RADARs use a relatively wide beam, it is very difficult to distinguish the structural details of the tracked object, but the distance measurement is relatively precise. In addition, RADARs can provide information about the target's velocity due to the Doppler effect. LiDARs are also active sensors that emit pulsed lasers to measure the range. Accordingly, they have very narrow beams with high resolution and greatly accurate distance measurement. However, LiDARs are susceptible to weather phenomena, like for instance, precipitation \citep{Thombre2022}. In such cases, the light may be broken or reflected and the sensors might provide misinformation. Despite that, they are utilizable at any time of the day and the high resolution offers the possibility to reconstruct the contours of the target. As discussed, every sensor type has its strengths and weaknesses. Some sensors are not working at night, some have problems with bad weather conditions, and some are not capable to classify objects. Therefore, the use of only one sensor is often not sufficient. In such a case, sensor fusion allows combining the benefits of different sensor types while eliminating their individual weaknesses. Fig.~\ref{fig:Sensors_fusion} demonstrates the basic idea of the sensor fusion concept.
Usually, uncertainty is an undesired quantity, but uncertainty is the key to sensor fusion. A more accurate estimation of the real state can be extracted using the intersection of several sensor uncertainties. Very often sensors are fused using Kalman filters \citep{Olfati_Saber2005cff, Olfati_Saber2007dkf}. However, RADAR sensors provide information about the distance of the tracked object, its related angle, and the distance change, which can be expressed in a radial velocity. Hence, the measurement model of a RADAR is nonlinear. A standard Kalman filter is not able to fuse RADAR signals. Here, nonlinear Kalman filters such as extended Kalman filters (EKFs) or unscented Kalman filters (UKFs) are more suited \citep{Chen2012kff}. EKFs approximate the model by linearizing it around the operating point, whereas UKFs propagate a set of sigma points through the nonlinear functions to better capture system dynamics. There are three primary sensor fusion strategies: early fusion, mid-level fusion, and late fusion. Early fusion combines raw sensor signals and learns a predictor by processing all the data together. Late fusion first processes each sensor individually, extracting information before combining their predictions. Mid-level fusion creates intermediate representations from the sensor data before learning a predictor. However, using a single-model estimation filter, such as for tracking a target with constant velocity, can lead to significant errors if the target's velocity changes. In such cases, interacting multiple model (IMM) filters \citep{Mazor1998} offer a more robust approach by connecting different model filters through a switching matrix that adjusts their weights according to the situation. While various sensor fusion techniques offer distinct advantages in combining data from multiple sources, the increasing complexity of sensor environments necessitates more sophisticated approaches. Factor graphs are a powerful tool for sensor fusion, representing complex relationships between variables as a bipartite graph of variable nodes and factor nodes. This structure allows for efficient encoding of multi-sensor information, enabling the integration of diverse data sources while maintaining a clear mathematical representation of dependencies \citep{Kschischang2001fga, Han2018aos, Zhang2023ams}. 

\subsubsection{Situational Awareness}
\label{sec: SITAW}
\begin{figure*}[ht]
	\centering
	\includegraphics[width=\textwidth]{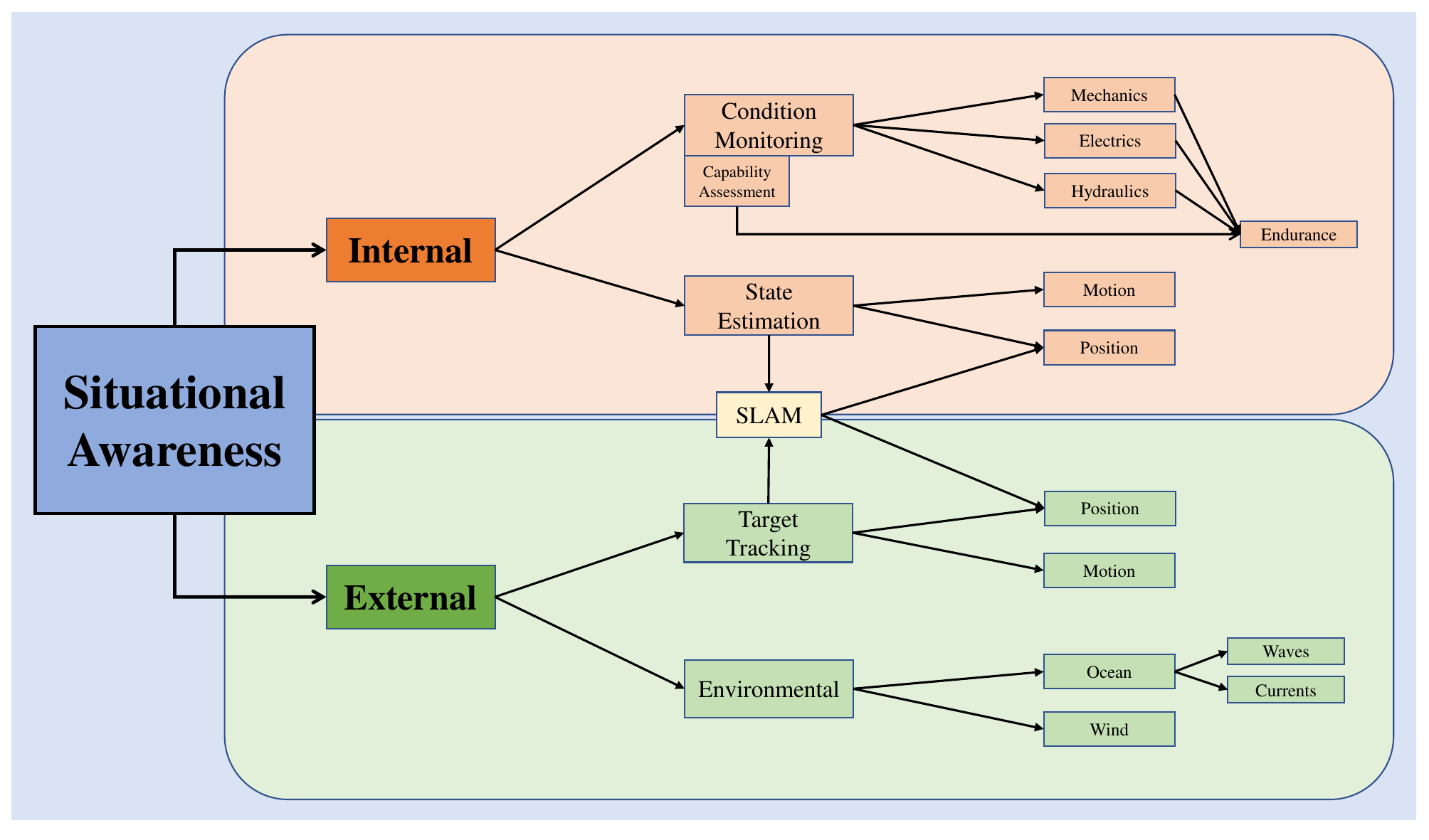}
	\caption{A basic overview of a vessel's situational awareness.}
	\label{fig:SITAW}
\end{figure*}

A comprehensive understanding of the vessel's condition and its environment is fundamental to developing an advanced digital twin. In general, the situational awareness (SITAW) of an autonomous vessel can be divided into internal and external components. The internal aspect covers self-awareness, such as condition monitoring and state estimation, while the external part involves target tracking and environmental awareness. These two areas are interdependent. For example, techniques like simultaneous localization and mapping (SLAM) can estimate both internal and external factors by mapping surrounding objects while estimating the ASV's motion and position. Although SITAW encompasses many more detailed components, the core areas are illustrated in Fig.~\ref{fig:SITAW} for simplicity.

\paragraph{Target Tracking}
Several studies are focusing on target tracking approaches of unmanned surface vehicles (USVs) using various types of detection sensors. The most common are RADARs, LiDARs, and cameras. To localize the global position, there is additionally the possibility to use GNSS and AIS, but AIS is often unreliable and reports in many cases incorrect vessel information \citep{Han2021}. However, by applying a UKF with AIS measurements, vessels can be tracked in geodetic coordinates over long distances \citep{Cole2021}. 
Various methods are proposed to improve trustworthiness, especially for short distances. By fusing You Only Look Once (YOLO) with scale-invariant feature transform (SWIFT), vessels are tracked based on camera data, deep learning and multiple features \citep{Zhang2020}. YOLO is the state-of-the-art camera classification algorithm capable of identifying objects in real-time. The basic idea of this approach is to split an image into bounding boxes and compute the probability that a box contains an object. Subsequently, the class probability is calculated, where usually a Convolutional Neural Network (CNN) is trained for. Finally, an intersection over union (IOU) of the dedicated class boxes is made, where all non-maximum IOUs are suppressed.
Furthermore, a tracking approach from parameterized LiDAR data is proposed, which uses random sample consensus (RANSAC) and continuously  adaptive mean shift (CAMShift) for boxing and centering objects \citep{Xing2019}. 
Via RADAR, vessel detection based on a recurrent neural network \citep{Jie2019}, and a belief rule-based methodology by utilizing particle filter is performed \citep{Liu2014}. However, vessel detection at short-range using RADARs is challenging due to their inherent shadow zone. Therefore, a vessel tracking approach fusing a pulse RADAR and a 3D LiDAR is proposed by \cite{Han2017}, where the motion is estimated by using an EKF. Other methods combine the estimation of the kinematic parameters using RADAR by additionally estimating the geometric parameters \citep{Han2021}. Recently, a new methodology called extended object tracking (EOT) received more and more attention by using ellipsoidal contour models. Thereby, a proposal is to track single elliptical targets in clutters with LiDAR sensors, for instance by combining an EKF with a generalized probabilistic data association (GPDA) filter \citep{Ruud2018}. GPDAs are associating data points to a tracked object. In this case, the detections are associated to an ellipse, which can lead to interpretation problems. If a vessel is tracked from behind, a small ellipse could be interpreted. Hence, sensor fusion may be the solution by using the velocity estimation of RADAR for pose extraction and LiDAR data to reconstruct the exact ellipse via EOT. However, most approaches have their problems with tracking multiple targets simultaneously. To tackle this non-trivial problem, algorithms such as joint probabilistic data association (JPDA) and joint integrated probabilistic data association (JIPDA) offer a suitable foundation. Here, the data association does not commit to a single detection but to a weighted combination of all detections, where closer detections are weighted higher. Furthermore, \cite{Menges2024dto} introduces a multi-target tracking methodology that integrates LiDAR and AIS data using Kalman Filtering (KF). This approach employs Density-Based Spatial Clustering of Applications with Noise (DBSCAN) to cluster LiDAR point clouds and subsequently fits ellipses to these clusters to enhance multi-target tracking capabilities.

\paragraph{State Estimation}
To estimate the exact position and velocity of a vessel, Kalman filters offer a trivial, but well-performing approach \citep{Perera_oceanvessel}. However, every sensor has uncertainties and thus using only IMU data does not offer a sufficient solution for autonomous systems. Therefore, IMU sensors are mostly fused with GPS data to obtain the exact position of the vessel. Kalman filters provide the possibility to fuse these sensors together \citep{Caron2006}. Furthermore, many other approaches are proposed, such as using automotive dead-reckoning \citep{Perumal2020} or visual SLAM \citep{Motlagh2019}. Moreover, several particle filter implementations are performed for state and parameter estimation. The most popular is the bootstrap filter, which draws the particles directly from stochastic equations. These characteristics enable a straightforward implementation approach but it could be problematic if many particles have to be generated. This constraint may be critical regarding real-time applications \citep{Chung2021}. Hence, the fusion of a UKF and a particle filter is suggested, where the degradation degree of the particle filter is reduced, but also the tracking precision is improved \citep{Xu2021}. The resampling of particle filters is one of the key components regarding precision and real-time execution. Therefore, a fast resampling scheme for particle filters is proposed by \cite{Li2013}, which provides comparable estimation accuracy.

\paragraph{Condition Monitoring}
Condition monitoring primarily falls under the scope of diagnostic DTs, which is discussed in detail in the following section. However, due to its relevance to a vessel's SITAW, it is briefly introduced here. To ensure the safe operation of a vessel's components, their conditions must be measured and verified, which is where condition monitoring becomes essential. This process provides critical information on mechanical, electrical, and hydraulic systems, among others, depending on the application. While advanced models for the mechanical and hydraulic systems of marine diesel engines already exist \citep{Nahim2015}, not all system states and parameters are easily measurable. As a result, observer design and parameter identification play key roles in modern condition monitoring. These techniques enhance the understanding of internal conditions by estimating hard-to-measure states. For example, while engine speed is straightforward to measure, its time derivatives are not, making adaptive observer methods useful for estimating unknown disturbances \citep{Xie2018}. Additionally, intake leakage detection observers are proposed by \cite{Ceccarelli2009} for evaluating system capability over extended periods.

\paragraph{Estimation of the Environment}
In the context of autonomous vessels, environmental factors can be categorized into ocean state and wind state. Accurately estimating the frequency and magnitude of ocean waves poses significant challenges. Various methods exist for determining wave characteristics, such as sea state estimation using multiple vessels acting as wave buoys \citep{Nielsen2019} or data-driven approaches that analyze motion responses from vessels \citep{Han2021a}. The introduction of the fast Fourier transform (FFT) algorithm has enabled real-time wave modeling by decomposing the discrete transformation matrix into sparse submatrices. While the estimation of sea currents remains underexplored in the context of MASS, some methods have been proposed in relation to underwater vessels \citep{Martinez2015, Hegrenaes2011}. Wind measurements are often compromised by the disturbance caused by a vessel's hull, leading to inaccuracies. One method to enhance measurement accuracy involves fusing data from multiple anemometers \citep{Rahmstorf1989}. Additionally, a three-axis anemometer has been suggested for independently measuring wind speed along each axis \citep{Gavrikov2021}. LiDAR measurements have also been utilized for wind field reconstruction, introducing a reduced-order model (ROM) of a higher-order dynamic wind model, with a UKF employed to estimate wind states \citep{Towers2014}. However, accurately modeling wind fields in real-time remains challenging due to hull-induced disturbances.
Considering the combined impact of wind, waves, and sea currents, the motion of a vessel has been described while simplifying the influences of these factors by \cite{Wang2022}. From a control-theoretical standpoint, observers can estimate unknown environmental disturbances affecting a vessel. An environmental disturbance observer framework has been proposed to estimate the forces impacted by wind, waves, and sea currents, accounting for model and measurement uncertainties by \cite{Menges2023aeda}.

\subsubsection{Modeling}
Modeling plays a crucial role in the functionality of descriptive digital twins since it helps bridge the gap between raw data and actionable insights. In the context of autonomous vessels, models serve several key purposes, including improving spatio-temporal resolution, denoising signals, correcting corruptions, addressing missing data, compressing data for efficient processing, and calculating critical quantities of interest that may be difficult to measure directly. Without robust models, the ability to interpret complex, multimodal data would be severely limited, hindering the overall performance of the digital twin.

The requirements for these models are stringent, especially since they often need to operate on edge devices embedded within the asset itself. First and foremost, the models must be computationally and memory efficient to ensure they can run in real-time without overloading the system. They must also be highly accurate and reliable since the critical decisions made by the autonomous system depend on the quality of the model's output. Certainty in the model’s predictions is vital, given the safety-critical nature of autonomous operations.

Additionally, explainability is a key requirement. The output of these models must be transparent and understandable to ensure that the decisions they inform can be trusted by human operators and regulatory bodies. Finally, the models must be generalizable and capable of handling a diverse array of scenarios, from routine operations to rare and unexpected events. Without these qualities, the effectiveness of descriptive digital twins in autonomous vessels would be severely compromised.

\begin{figure*}[t!]
	\centering
	\includegraphics[width=\textwidth]{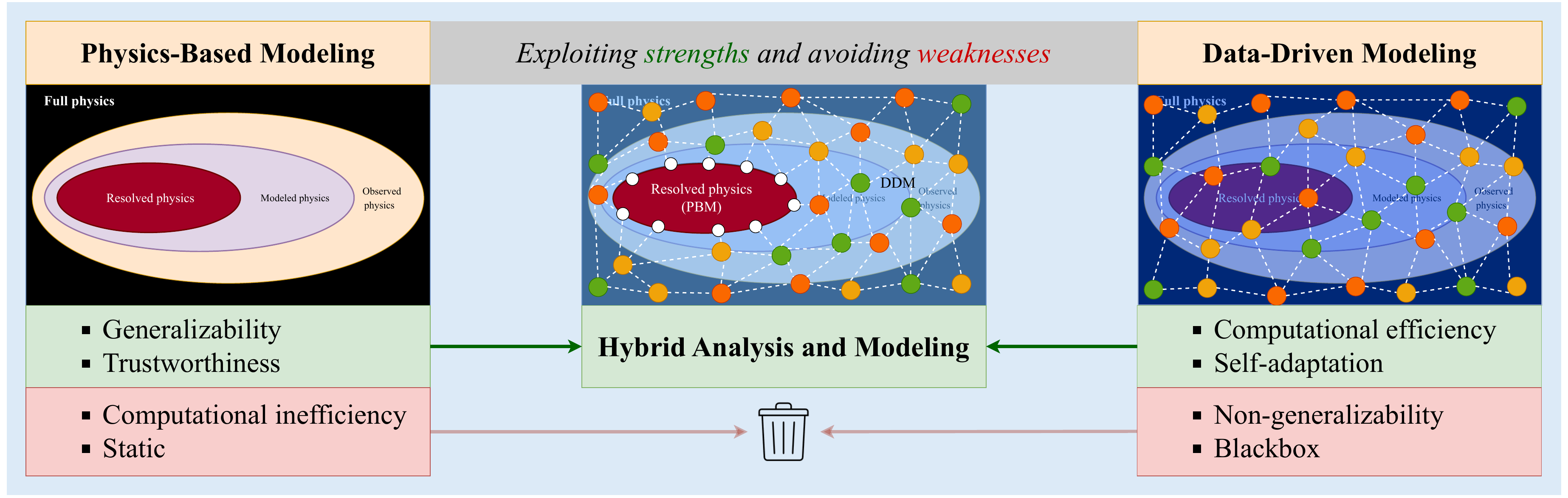}
	\caption{The concept of Hybrid Analysis and Modeling (HAM).}
	\label{fig:HAM}
\end{figure*}

\paragraph{Physics-based modeling}
Physics-Based Modeling (PBM) (Fig.~\ref{fig:HAM}) is grounded in the fundamental principles of physics, such as fluid dynamics, thermodynamics, and structural mechanics, to model the behavior of autonomous surface vessels. PBM utilizes mathematical equations to describe interactions between the vessel and its environment, such as wave impact, resistance, and propulsion forces. These models are highly accurate in predicting the vessel's behavior under known conditions because they rely on well-established physical laws. However, PBMs are often computationally intensive, especially when considering complex systems or dynamic environments. As a result, real-time applications can be challenging, and these models may not be flexible enough to handle unexpected scenarios or unmodeled phenomena that could arise during vessel operations.

\paragraph{Data-driven modeling}
Data-Driven Modeling (DDM) (Fig.~\ref{fig:HAM}), by contrast, does not rely on explicit physical laws but instead extracts patterns and relationships directly from large amounts of data collected from sensors or historical vessel operations. Using techniques such as machine learning, DDM can quickly identify trends and predict vessel behavior under various conditions. Data-driven models are highly adaptable, capable of adjusting to new or unforeseen situations without needing an in-depth understanding of the underlying physical processes. This adaptability makes DDM especially useful in real-time applications, where computational efficiency is critical. However, DDM often lacks transparency and interpretability, making it difficult to understand how decisions or predictions are made. Additionally, data-driven models can suffer from biases or inaccuracies if the training data is not representative of the actual operational environment.

\paragraph{Hybrid analysis and modeling}
Hybrid Analysis and Modeling (HAM) (Fig.~\ref{fig:HAM}) seeks to combine the best of both domains by integrating PBM with DDM techniques. In this approach, physical laws provide a foundation to ensure that the model remains grounded in reality, while data-driven elements allow for flexibility and real-time adaptability. This combination can lead to more accurate predictions than either method alone since HAM integrates the precision of physics-based models and the adaptability of data-driven models. The concept of HAM is demonstrated in Fig.~\ref{fig:HAM}. For autonomous surface vessels, this approach is particularly valuable because it enables real-time decision-making without sacrificing the accuracy needed for critical operations. Hybrid models can adapt to a wide range of conditions while maintaining computational efficiency, making them ideal for complex systems where both interpretability and adaptability are important.

\subsection{Level 2: Diagnostic}
Diagnostic models serve as crucial tools for assessing the condition and operational health of ship systems and machinery. By utilizing data from acoustic, temperature, and vibration sensors, these models enable early detection of anomalies and potential failures, facilitating proactive maintenance and reducing the risk of unexpected downtime. Each diagnostic method focuses on a specific aspect of the vessel's performance, from vibration and lubricant condition to temperature and structural health, ensuring comprehensive monitoring and enhancing the reliability of maritime operations.

\subsubsection{Condition Monitoring}
Generally, the area of condition monitoring can be divided into three classes: reactive maintenance, preventive maintenance, and predictive maintenance. Reactive maintenance only intervenes if a component fails completely. Preventive maintenance initiates and takes action in a predefined cycle to avoid critical outages. However, preventive maintenance can be very cost-intensive. Predictive maintenance prevents spurious actions based on probabilities describing the future. While predictive maintenance is widely used in other fields, most vessels contain at most preventive maintenance \citep{Gkerekos2017}. The estimation of endurance necessitates the inclusion of predictive elements. Especially for autonomous vessels it is important to early realize an outage of a component in order to premature substitute it.

\paragraph{Vibration Analysis Models} Vibration analysis models are employed to monitor and assess the dynamic behavior of ship machinery, including engines and propulsion systems, for early detection of irregularities and potential failures. Deviations from expected vibration patterns can reveal critical issues such as misalignment, bearing degradation, or mechanical imbalance \citep{Weddell2012vps, Rao2022aro, Mafla_Yepez2024ava}.

\paragraph{Lubricant Condition Monitoring Models} These models analyze the condition of lubricating oils to identify signs of wear, contamination, and degradation. Monitoring oil quality helps predict potential failures in engines and other machinery \citep{Wu2023aes, Li2024mol}. Given that lubricant is a critical component of a ship’s power plant, assessing changes in lubricant viscosity and the characteristics of wear particles is essential for effective condition monitoring. 

\paragraph{Temperature Monitoring}
Infrared thermography is employed to monitor the temperatures of critical components within the system. The identification of hot spots or abnormal temperature distributions can signal underlying issues, such as electrical faults, overheating, or excessive friction in machinery \citep{Menges2024cam, Bagavathiappan2013itfa}. This non-invasive technique enhances predictive maintenance capabilities by facilitating early detection of potential failures \citep{Menges2024rtp}.

\paragraph{Structural Health Monitoring (SHM)} Hull integrity models evaluate the structural health of the ship's hull, detecting potential issues such as corrosion, cracks, and deformation. Continuous monitoring is essential for maintaining the vessel's structural integrity and ensuring its safe operation \citep{Shen2015aso, Hu2007anm}.

\paragraph{Acoustic Emission Models:} These models detect and analyze acoustic signals emitted by materials under stress \citep{Jiang2008mod, Anastasopoulos2009aem}. Acoustic monitoring is used for detecting structural defects, assessing the integrity of components, and identifying potential failures. Furthermore, this method provides a non-invasive means of evaluating structural health, allowing for continuous monitoring without disrupting operational processes.

\paragraph{Performance Monitoring}
Monitor the performance of engines and propulsion systems by tracking parameters such as fuel consumption, power output, and efficiency \citep{Gupta2023ssa,Gupta2021smo}. Deviations from expected performance \citep{Gupta2024cbo} can signal mechanical issues or inefficiencies in operation. Additionally, this monitoring enables timely interventions, optimizing maintenance schedules and improving overall operational reliability \citep{nikula_data-driven_2016, Perera2016daf, Perera2016mec}.

\paragraph{Remote Monitoring and Internet of Things (IoT)}
Internet of Things (IoT) refers to a network of interconnected devices that collect and exchange data. Deploying IoT sensors throughout the vessel to monitor various parameters such as temperature, pressure, and vibration. Data collected from these sensors is transmitted in real-time for analysis, enabling proactive maintenance and early fault detection \citep{Milic2020ttf, Salah2020ies}.

\paragraph{Corrosion Monitoring}
It involves predicting and monitoring the rate of corrosion on the ship's structure \citep{Panayotova2007cmo}. Understanding corrosion patterns helps in implementing preventive measures and scheduling maintenance activities \citep{Agarwala2000cda}. Condition monitoring models in the shipping and maritime industry aim to enhance reliability, prevent unplanned downtime, and optimize maintenance strategies. The integration of these models allows for a more comprehensive understanding of the vessel's health and performance.

\paragraph{Data Analytics and Machine Learning}
Machine learning algorithms utilize advanced data analytics to process and interpret vast datasets collected from diverse sensor systems. These models excel at detecting intricate patterns, anomalies, and potential failure modes \citep{Gupta2022spm} that are often difficult to discern through traditional analytical methods. A proposed method for diesel engine anomaly detection uses auto-associative kernel regression (AAKR) with a memory matrix trained on historical fault-free data to detect deviations in operation \citep{Wei2016}. While it requires data from 46 sensors (e.g., rotation, temperature, pressure, vibration), other studies show thermal parameters alone, analyzed via Mahalanobis distances, can effectively detect faults in marine diesel engine valve systems \citep{Sun2020}.

\subsubsection{Cybersecurity} \label{sec: Cybersecurity}
The integration of cybersecurity and risk management gets more and more obligatory in commercial shipping. This holds especially for autonomous vessels, where the importance of securing the technical components from potential manipulation attacks increases vastly. In \cite{Lee2023}, the vulnerability of machine learning methods, especially YOLOv5, to adversarial attacks for enhanced cybersecurity in MASS is evaluated. It is concluded that only small perturbations directed to the gradient of the loss function can affect a DNN, making the interpretations and classifications useless. At the same time, a human observer would not see a difference between the original and the perturbed image. Therefore, \cite{Yoo2023} formulates cybersecurity requirements for autonomous vessels, focusing on sensor data, artificial intelligence (AI), and how potential threats can affect a vessel's SITAW, navigation, and control. In parallel, institutions such as the international maritime organization (IMO) try to define global guidance and standards. According to the IMO's guidelines on cybersecurity, cyber incidents can arise as the result of, for instance \cite{InternationalMaritimeOrganization2}:
\begin{itemize}
	\item A cyber security incident, which affects the availability and integrity of operational technology, for example corruption of chart data.
	\item An unintended system failure occurring during software maintenance and patching, for example
	through the use of an infected USB drive to complete the maintenance.
	\item Loss of or manipulation of external sensor data, critical for the operation of a ship. This includes
	but is not limited to  GNSS, of which the GPS is the most frequently used.
	\item Failure of a system due to software crashes and/or “bugs”
	\item Crew interaction with phishing attempts, which is the most common attack vector by threat
	actors, which could lead to the loss of sensitive data and the introduction of malware to
	shipboard systems.
\end{itemize}
In this context, the method of cybersecurity protocols will continue to evolve, while the used cybersecurity concepts will be mostly similar and require frequent software patch upgrades. In the future, these issues might be solved by itself if a transition from classical approaches to quantum encryption and quantum communication will happen \citep{McGillivary2018}. Quantum communication combines classical information theory and the theory of quantum mechanics. By using quantum states to carry information, the field of quantum cryptography is expected to break through the limits of classical communication and encryption \citep{Zhang2019}. Either way, the importance of cybersecurity regarding autonomous surface vessels is undisputed and has to be addressed in the future.

\subsection{Level 3: Predictive}
In autonomous shipping, a predictive digital twin plays a crucial role in online operational management rather than offline planning. It continuously monitors real-time data from the vessel’s systems, such as engines, navigation, and sensors, to make immediate adjustments and optimize performance during operations. By predicting potential system failures and adjusting to changing sea conditions in real-time, it ensures uninterrupted, efficient, and safe autonomous navigation. This live operational insight allows the vessel to respond dynamically to situations, improving decision-making and minimizing downtime while the ship is actively in service.

\paragraph{Voyage Time Prediction Models} These models use historical data, weather forecasts \citep{Scheuerer2015pws, Zyczkowski2019row, Zhang2021aro}, and real-time conditions to predict the estimated time of arrival (ETA) for vessels. Accurate voyage time predictions assist in better planning for port operations, resource allocation, and overall logistics \citep{Lu2015ase, Yoon2023ecv, Magalhaes2023vvt, Zhang2024poc, Schindler2024tva}. Furthermore, integrating predictions of the sea state into these models enhances the accuracy of ETA calculations by accounting for the potential impacts of waves, currents, and other maritime conditions.

\paragraph{Fuel Consumption Prediction Models} Predictive models that estimate fuel consumption based on factors such as vessel speed, route, weather conditions, and engine efficiency. These models help optimize fuel usage, reduce costs, and improve environmental sustainability \citep{BalBesikci2016aan, Wang2018psf, Hu2019pof, Li2019pos, Yan2020doa, Yan2021daf}. Incorporating real-time weather forecasts allows for more precise fuel consumption estimates since adverse weather can significantly affect vessel performance. Additionally, predicting the sea state aids in identifying optimal routes that minimize fuel usage.

\paragraph{Emissions Prediction Models} Models that predict the emission output of vessels based on operational parameters \citep{Jalkanen2009ams, Lu2015ase, Larsen2015doa, Yan2023cfp}. This supports compliance with environmental regulations and facilitates the implementation of sustainable shipping practices. By factoring in variables such as fuel type, speed, and engine load, these models can provide insights into how different operational choices influence emissions, aiding in the development of greener shipping practices.

\paragraph{Equipment Failure Prediction Models} Using machine learning and condition monitoring data, these models predict the likelihood of equipment failures. By identifying potential issues before they occur, maintenance can be scheduled proactively, minimizing downtime and reducing repair costs \citep{Goksu2020pos, Kim2023uer, Menges2024rtp}. Additionally, predicting the life endurance of ship components enhances maintenance planning by allowing operators to replace or service parts before they fail, thereby improving overall operational reliability. 

\paragraph{Weather Routing Models} Predictive models that consider real-time and forecasted weather conditions to optimize ship routes for fuel efficiency and safety. These models help vessels navigate around adverse weather, reducing fuel consumption and enhancing crew safety \citep{Scheuerer2015pws, Zyczkowski2019row, Zhang2021aro}. By incorporating detailed predictions of the sea state, these models can better assess the risks associated with specific routes, allowing for adjustments that ensure both safety and efficiency.

\paragraph{Cargo Arrival Time Prediction Models} These models predict the arrival times of cargo shipments, allowing ports, logistics companies, and other stakeholders to plan and coordinate activities more effectively \citep{Yoon2023ecv, Zhang2024poc, Schindler2024tva, Postan2019doa}. Accurate predictions of cargo arrival times also depend on effective tracking of target trajectories, which allows for adjustments in logistics and resource management.

\paragraph{Demand Forecasting Models} Forecasting models that predict future cargo demand based on historical data, market trends, and economic indicators. Accurate demand forecasting supports efficient resource allocation and planning for shipping companies \citep{Meng2015atb, Patil2017sdd, ElNoshokaty2019tio}. Predictive models that forecast the demand for container space can optimize container loading strategies. This helps maximize the use of available cargo space and improve overall operational efficiency. Incorporating predictions based on economic indicators enhances the accuracy of demand forecasts, enabling better planning and responsiveness to market changes.

\paragraph{Port Congestion Prediction Models} Using historical and real-time data, these models predict port congestion, enabling shipping companies to adjust routes or schedules to avoid delays and minimize disruptions \citep{Zhang2024poc, Patil2017sdd, Tian2023asp, Pruyn2020aop, Peng2023adl}. Predicting demand for port services alongside congestion trends allows for more strategic planning, helping to alleviate bottlenecks.

\paragraph{Crew Management Models} Predictive models that assist in forecasting crew availability, fatigue levels, and training needs. Effective crew management enhances safety and reliability in maritime operations by ensuring that personnel are well-prepared and capable of responding to various operational demands \citep{Ung2015awc, Grech2009aeo, Maceiras2024aom, Heij2018ppo}. By incorporating predictions of crew fatigue based on workload and operational schedules, shipping companies can better manage personnel resources to maintain safety and performance standards.

\paragraph{Market Trend Analysis Models} Analytical models that assess market trends, shipping rates, and geopolitical factors to help shipping companies make informed decisions about vessel deployment, route planning, and long-term business strategies \citep{Maceiras2024aom, Heij2018ppo}. These models can benefit from real-time data on global economic indicators, enhancing the accuracy of market forecasts and strategic planning.

\paragraph{Cybersecurity Threat Prediction Models} Predictive models that analyze cybersecurity threats and vulnerabilities to anticipate potential cyberattacks on maritime systems. This helps enhance the cybersecurity posture of shipping companies \citep{Kelemen2018cit, Guo2023rtr, Antonopoulos2022dap}. By considering the evolving nature of cyber threats and integrating predictive analytics, these models support proactive measures to safeguard critical maritime operations.

\subsection{Level 4: Prescriptive}
At Level 4, prescriptive models emerge as a sophisticated extension of the predictive models developed in Level 3. These predictive models lay the foundation for exploring a variety of operational scenarios using historical data and real-time insights. By accurately forecasting vessel performance, fuel consumption, and potential risks, predictive models provide a rich context for decision-making. They enable stakeholders to simulate different situations, such as varying weather conditions or unexpected port delays, allowing for a comprehensive understanding of how an autonomous surface vessel might respond in each scenario.

Transitioning to Level 5, the role of autonomous models becomes critical for enhancing the effectiveness of prescriptive approaches. While prescriptive models outline the best courses of action based on predictive insights, they often require intelligent interventions from controllers to adapt to dynamic environments. These intelligent actions are vital for accurately predicting the outcomes of various operational decisions, ensuring that vessels can respond effectively to real-time changes. Thus, the synergy between predictive, prescriptive, and autonomous models not only optimizes operational strategies, but also improves the resilience and adaptability of autonomous surface vessels in complex maritime landscapes.

\paragraph{Optimizing Operational Efficiency through Prescriptive Analytics}
Prescriptive analytics employs data and advanced algorithms to recommend optimal actions for autonomous surface vessels, enhancing operational efficiency and safety. By integrating real-time data from the vessel’s digital twin, these models provide actionable insights that guide decision-making in various operational scenarios.

\paragraph{Route Optimization for Enhanced Safety and Efficiency}
One key application is in route optimization. By considering factors such as fuel consumption, weather conditions, and maritime traffic, prescriptive models can suggest the most efficient routes that minimize travel time and costs while ensuring safety. For instance, a digital twin can analyze historical data and real-time environmental inputs to recommend route adjustments that avoid adverse weather, thus enhancing both safety and fuel efficiency.

\paragraph{Proactive Maintenance Scheduling}
Another critical aspect is maintenance scheduling. By assessing the condition and life expectancy of various components, prescriptive models can suggest optimal maintenance windows that minimize downtime while maximizing the vessel's operational readiness. This proactive approach not only reduces the likelihood of unexpected failures but also extends the lifespan of critical machinery.

\paragraph{Enhancing Crew Management Strategies}
Furthermore, prescriptive analytics can play a significant role in crew management. By evaluating crew availability, fatigue levels, and training needs, these models can recommend staffing levels and training schedules that ensure the crew is well-prepared for their operational responsibilities. This is especially important in autonomous operations, where the human element must be optimized to support automated systems effectively.

\paragraph{Ensuring Regulatory Compliance through Simulation}
In addition, prescriptive models can aid in compliance with regulatory requirements by suggesting operational adjustments that meet safety and environmental standards. For example, a digital twin can simulate various operational scenarios and recommend practices that align with the latest regulations, thus reducing the risk of non-compliance.

\paragraph{Advancing Maritime Industry Resilience and Sustainability}
Overall, the integration of prescriptive analytics into the digital twin framework for autonomous surface vessels not only enhances operational decision-making but also contributes to a more resilient and adaptive maritime industry. Using these advanced models, operators can achieve a competitive advantage while promoting sustainability and safety in maritime operations.

\subsection{Level 5: Autonomous}
\begin{figure*}[ht]
	\centering
	\includegraphics[width=\linewidth]{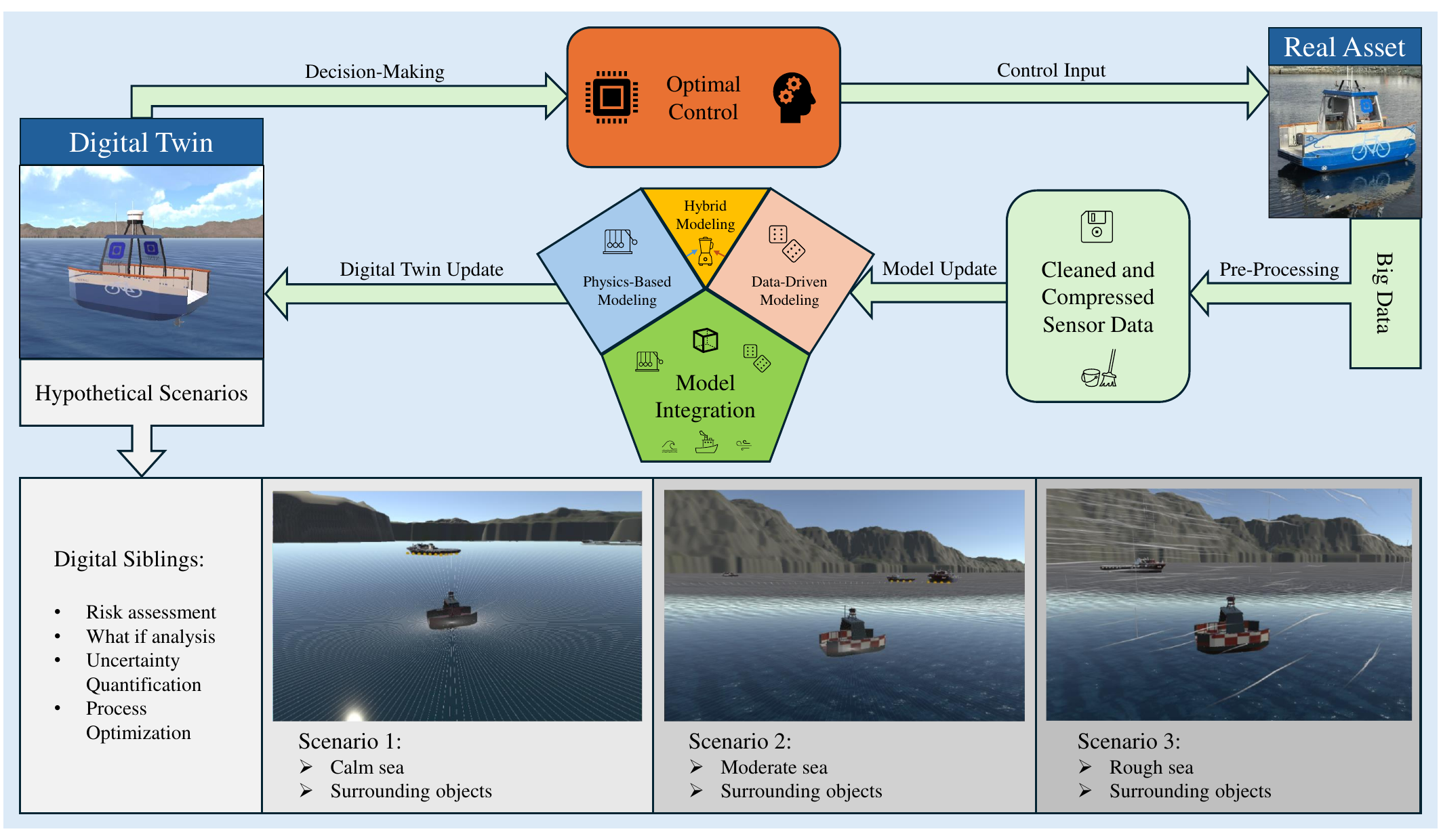}
	\caption{General principle of an autonomous digital twin. Green arrows represent real-time updates, while grey arrows indicate offline simulations.}
	\label{fig:smart_DT}
\end{figure*}

Fig.~\ref{fig:smart_DT} illustrates a conceptual framework for an autonomous digital twin, demonstrating how such a system could operate. The green arrows indicate real-time connections between the physical vessel and its digital counterpart, while the grey arrows represent offline processes, such as simulations and data analysis. Key components of this autonomous DT include optimal control, ensuring real-time decision-making, modeling for physical realism, which guarantees accurate representation of the vessel's behavior, and hypothetical scenarios, allowing for the exploration of different outcomes. Other features, such as digital siblings for scenario testing and big data integration for improved forecasting and decision-making, highlight the advanced capabilities that such a system could offer.

\subsubsection{Path Following and Collision Avoidance} \label{sec: COLAV}
Coordinating shipping traffic autonomously is a challenging task due to the complexity of maritime environments. While the COLREG (see Section~\ref{sec: Standardization}) provides clear guidelines, many real-world scenarios are difficult to interpret and implement precisely. This makes the development of robust and reliable decision-making algorithms essential but complicated. Additionally, human operators occasionally make mistakes, necessitating that autonomous vessels are capable of responding quickly and safely.

Various algorithms have been proposed to tackle path following and collision avoidance, including colony algorithms, genetic algorithms, fuzzy logic, and rapidly exploring random tree (RRT) algorithms. However, many of these struggle with real-time route optimization. For example, artificial potential field (APF) methods can help manage obstacle avoidance \citep{Huang2020}, but they may create local minima, requiring supplementary algorithms to ensure stable performance. Alternative approaches, such as game theory \citep{Lisowski2012} and optimal control \citep{Miele1999}, offer additional decision-making frameworks.

Recent advancements have introduced solutions such as model predictive control (MPC) for path following and collision avoidance \citep{Hagen2018, Du2021, Papadimitrakis2021}, while reinforcement learning (RL) has also emerged as a popular approach in this domain \citep{Meyer2020taa, Meyer2020ccc}. A hybrid approach that combines RL and MPC uses Q-learning to optimize parameters for enhanced closed-loop performance, which are then applied to the MPC cost functions \citep{Martinsen2022}. Most collision avoidance research focuses solely on avoiding obstacles, but few studies address both target tracking and collision avoidance. However, in some studies \cite{Kufoalor2019} integrate radar-based tracking with MPC, aligning with COLREG to ensure compliance and safety. To fully realize autonomous vessel operations, robust control mechanisms are crucial. In the following section, the state-of-the-art controllers employed in autonomous surface vessels are elaborated.

\subsubsection{Control Techniques}
\paragraph{Proportional-Integral-Derivative (PID) Control}
PID controllers are widely used for basic control tasks, such as maintaining a desired heading, speed, or depth \citep{Hu2013tcc, Dewi2021dco, Nan2020ddr, Feng2021dar}. They are effective in stabilizing the vessel and minimizing deviations from the setpoint. Simplified versions of PID controllers, such as the proportional-only (P) controller, also exist. For instance, in \cite{Kinaci2023sdt}, a P controller is used with a digital twin to validate a maneuvering model and controller algorithms through simulations. Many PID-based control methods are enhanced with advanced techniques like fuzzy logic \citep{Yunsheng2015ofs, Majid2015afs} or combined with neural networks \citep{Fang2017acc}.

\paragraph{LQR (Linear Quadratic Regulator) Control}
LQR controllers are designed for linear systems and are effective in stabilizing the vessel by minimizing a quadratic cost function. They are often used for linearized models of ship dynamics \citep{Seo2022lsa, Feng2021dar, Yin2024sab}.

\paragraph{Backstepping Control}
Backstepping control is a recursive design method used for stabilizing nonlinear systems, making it suitable for the complex dynamics of autonomous surface vessels. It is often applied to tasks like heading and trajectory control, ensuring stable and accurate maneuvering \citep{Kinjo2022bco, Nan2020ddr, Jin2018dav, Abrougui2023adf, Wang2021fco, Piao2020abs}.

\paragraph{Sliding Mode Control}
Sliding mode control is a robust technique for handling nonlinear systems and uncertainties, frequently applied to autonomous surface vessels to maintain stability in challenging conditions. It ensures effective performance even in the presence of disturbances such as waves and wind \citep{Nan2020ddr, Piao2020abs, Wan2019asm, Zhu2024ttc, Kawamura2019co, Liu2023mfa}.
    
\paragraph{Adaptive Control}
Adaptive control algorithms adjust the control parameters based on changes in the vessel's dynamics or environmental conditions. This adaptability is particularly useful in handling uncertainties \citep{Baek2022mra, Jiang2019hco, Liu2015lpv, Liao2020rom}.

\paragraph{Model Predictive Control (MPC)}
MPC is a sophisticated control strategy that uses a predictive model of the system to optimize control inputs over a specified prediction horizon. MPC is suitable for complex tasks like trajectory planning, obstacle avoidance, and energy optimization \citep{Veksler2016dpw, Hagen2018mbc, Zheng2020rdp, Han2022apf, Trym2020ocr, Menges2024nmp}. 

\paragraph{Machine Learning-Based Control}
Reinforcement learning and other machine learning approaches are increasingly used for control in maritime autonomy. These algorithms learn from experience and can adapt to changing environmental conditions and dynamic situations \citep{Moe2018mli, Meyer2020ccc, Meyer2020taa, Heiberg2022rbi, Larsen2021cdr}. However, this class of models is computationally demanding specially when the perception data is high dimensional. Another issue with these models is that there is blackbox deep neural network at the core learning the optimal policies. Recently the use of variation autoencoder has been demonstrated by \cite{Larsen2024vaf} to address the curse of dimensionality associated with the perception data. Moreover, \cite{Vaaler2024mca} demonsteated the use of safety filters to make this class of model safe for training on real vessels. 

\paragraph{Fuzzy Logic Control}
Fuzzy logic controllers are employed when dealing with uncertainties and imprecise information. They can be useful in situations where the relationship between input and output is not well-defined \citep{Xiang2018sof, Yunsheng2015ofs, Li2024cco}.
    
\paragraph{Swarm Control Algorithms}
In scenarios involving multiple autonomous vessels, swarm control algorithms enable coordinated and collaborative behavior. These algorithms allow vessels to communicate and work together to achieve common objectives \citep{Zhu2011ftc, Guo2020gpp, Krell2022asv, Wang2021lpo}.
    
\section{Conclusion and recommendations} \label{sec:conclusionsandrecommendations}
To conclude, enabling technologies for every capability level of digital twins, from standalone systems to fully autonomous operations, are already in existence with varying levels of maturity. These technologies, which include advancements in sensor integration, predictive analytics, real-time data processing, and advanced controllers, provide the necessary foundation for developing digital twins tailored to autonomous surface vessels. 

\subsection{Value and challenges}
\label{subsec:valuesandchallenges}
If fully realized, digital twins will enable significant improvements in the following broad areas:
\begin{itemize}
    \item \textit{Enhanced Safety:} One of the most significant benefits of employing digital twin technology in autonomous surface vessels (ASVs) is the enhancement of safety in maritime operations. By creating a virtual representation of the vessel and its operating environment, digital twins allow for real-time monitoring, predictive analytics, and advanced decision-making. This enables the identification of potential hazards, such as collisions, equipment malfunctions, or navigational errors, well before they pose a threat to the vessel or its crew. Additionally, the use of digital twins allows for the continuous testing of critical systems in a virtual environment, reducing the risks associated with human error and unpredictable sea conditions. With real-time data inputs from onboard sensors, the digital twin can predict and respond to dangerous situations, providing a proactive approach to maritime safety that is far superior to traditional reactive methods.
    \item \textit{Cost-Effective Operation:} Digital twin technology plays a crucial role in optimizing the cost-effectiveness of ASVs by streamlining operations and reducing downtime. Through advanced simulations and real-time data integration, digital twins enable predictive maintenance, which allows operators to anticipate component failures before they occur, minimizing expensive emergency repairs and avoiding unscheduled downtime. Additionally, by optimizing fuel consumption and routing based on real-time environmental data, ASVs can reduce operational costs, enhance fuel efficiency, and decrease wear and tear on the vessel. This dynamic optimization ensures that ASVs operate at peak performance under varying conditions, significantly cutting operational costs while maintaining high standards of service and reliability. Moreover, the digital twin’s ability to simulate different operational scenarios enables shipping companies to explore various strategies without incurring real-world costs, further driving cost savings.
    \item \textit{Environmental Sustainability:} Incorporating digital twin technology into ASVs also supports the push for more environmentally sustainable maritime operations. By enabling precise fuel management, digital twins can help minimize fuel consumption, which in turn reduces carbon dioxide (CO2) and other greenhouse gas emissions from shipping activities. The ability to simulate and optimize ship routes based on weather conditions and sea currents also results in reduced energy use, making shipping more efficient and less harmful to the environment. Additionally, digital twins allow for the monitoring and simulation of a vessel’s environmental impact, including emissions and waste management, helping maritime operators comply with increasingly stringent environmental regulations. As sustainability becomes a global priority, the implementation of digital twins in ASVs will be crucial in making the maritime industry greener and more responsible.
\end{itemize}

However, despite these technological advancements, the integration of these technologies into fully operational digital twins remains a challenge. The main stumbling blocks can be itemized as follows:

\begin{itemize}
    \item \textit{Data and Communication:} The seamless flow of data between the physical vessel and its digital twin is critical for real-time decision-making and autonomous operations. However, managing the high volume, velocity, variety, and veracity (the "4Vs" of big data) from diverse sensors poses significant challenges. Reliable data transmission in maritime environments, especially in remote areas or under harsh conditions, requires robust communication systems. This includes addressing latency issues, bandwidth limitations, and ensuring data integrity. Moreover, the integration of data from multiple heterogeneous sources (such as RADAR, LiDAR, cameras, and IoT devices) into a cohesive model requires advanced data fusion techniques and standardized communication protocols to ensure interoperability between systems.
    \item \textit{Security:} With the increasing reliance on data-driven systems and interconnected networks, cybersecurity is a major concern for autonomous vessels and their digital twins. Autonomous ships are vulnerable to cyberattacks that could manipulate data, disrupt communication, or even take control of the vessel. Ensuring secure data transmission, safeguarding onboard systems from external threats, and establishing robust authentication and encryption protocols are critical to maintaining operational integrity. The challenge is heightened by the complexity of integrating cybersecurity across a broad range of sensors, communication networks, and control systems, making it essential to address potential vulnerabilities from the design stage through to deployment and operation.
    \item \textit{Modeling:} Accurate and computationally efficient models are the backbone of any digital twin, as they are essential for simulating real-world behaviors and predicting future states. However, developing models that can capture the full range of a vessel’s operational conditions, including hydrodynamics, mechanical performance, and environmental influences, is a complex task. These models must balance precision with the need for real-time processing, especially in dynamic and unpredictable maritime environments. Moreover, combining data-driven models with physics-based simulations in a way that remains transparent, explainable, and reliable is still an ongoing research challenge, particularly in scaling up from simple systems to the complexity of fully autonomous vessels.
    \item \textit{Regulations:} Regulatory frameworks for autonomous surface vessels and digital twins are still evolving, and there are currently no widely accepted international standards governing their development and deployment. Maritime laws such as the International Regulations for Preventing Collisions at Sea (COLREG) are primarily designed for human-operated vessels, and adapting these rules to autonomous operations presents legal and ethical dilemmas. Governments and international bodies, like the International Maritime Organization (IMO), are working towards developing comprehensive guidelines for Maritime Autonomous Surface Ships (MASS), but regulatory uncertainty remains a significant barrier. Compliance with regional and international regulations, as well as safety certifications, will be necessary before digital twins can be fully integrated into operational maritime systems.
    \item \textit{Industrial Acceptance:} Gaining industrial acceptance for digital twins and autonomous vessels is another key challenge. Shipowners and operators may be hesitant to adopt these technologies due to concerns about the reliability of autonomous systems, the cost of implementation, and the potential disruption to established workflows. Convincing stakeholders of the tangible benefits, such as reduced operational costs, enhanced safety, and improved environmental performance, will be essential. Additionally, training personnel to interact with and trust these advanced systems, as well as integrating digital twins into existing maritime operations, will require significant time, investment, and changes to the industry’s mindset.
\end{itemize}

By addressing these opportunities and challenges, stakeholders can work toward the successful implementation of highly capable digital twins for autonomous surface vessels, ultimately improving the safety and efficiency of maritime operations.

\begin{figure*}[ht!]
\centering
\includegraphics[width=\linewidth]{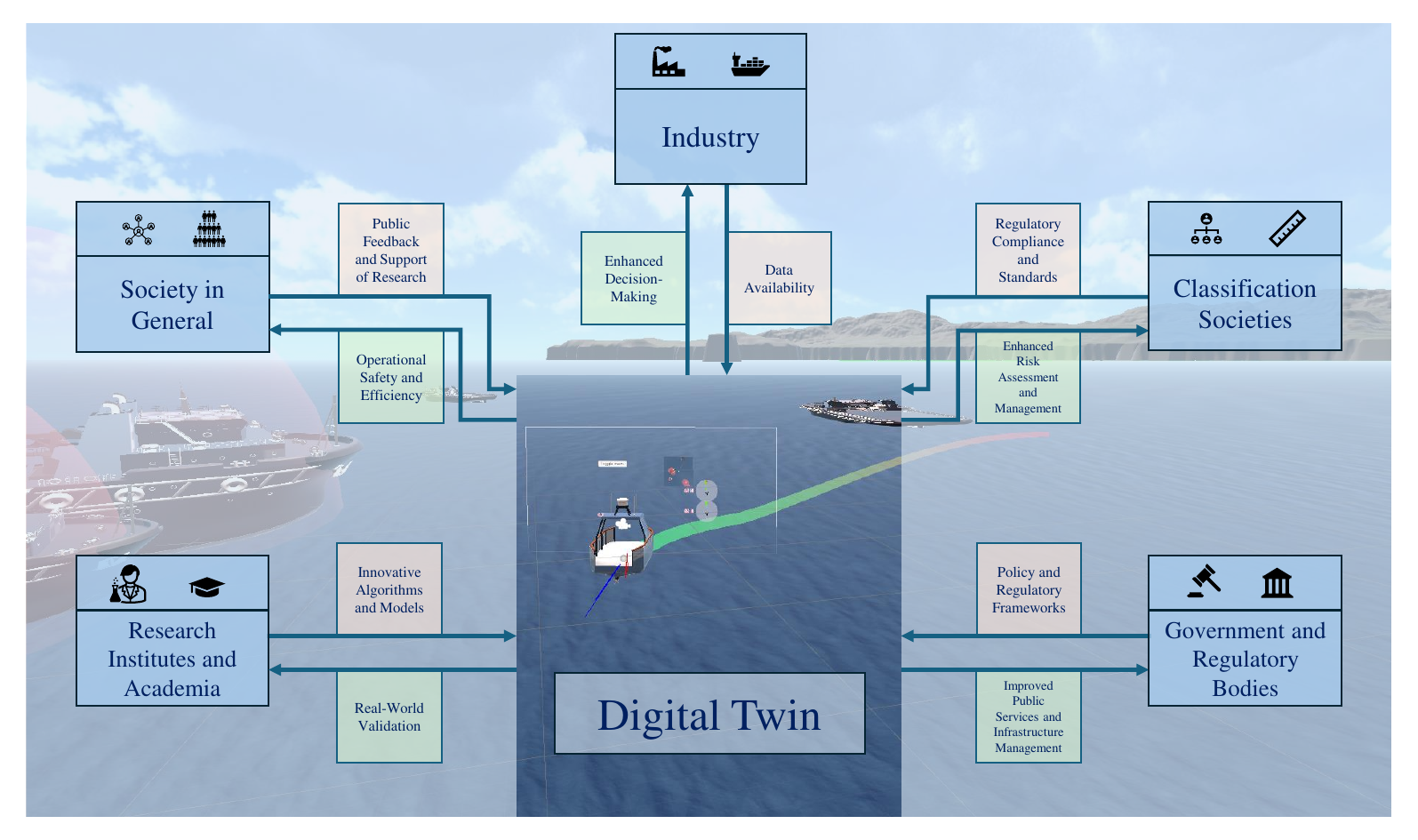}
\caption{Stakeholders and their potential contributions.}
\label{fig:stakeholders}
\end{figure*}

\subsection{Stakeholder Roles and Recommendations}
Based on the context of Digital Twins for Autonomous Surface Vessels and the identified knowledge gaps, the following recommendations are proposed for different stakeholders. Fig.~\ref{fig:stakeholders} illustrates the key stakeholders, their expected contributions, and the positive impact that digital twins can have on their operations.

\subsubsection{Industry (Maritime and Technology Companies)}
\textbf{Role}: Develop and implement digital twin technologies by collaborating with sensor manufacturers, software developers, and maritime operators. Industry stakeholders should prioritize integrating Digital Twin systems with real-time data and advanced analytics for condition monitoring, optimization, and autonomous operations.

\textbf{Recommendations}:
\begin{itemize}
    \item Collaborate with research institutes and government bodies to define standards for data exchange between physical assets and their digital twins.
    \item Invest in big data analytics, sensor fusion, and hybrid modeling techniques to improve decision-making and risk assessments for autonomous vessels.
    \item Lead the way in testing and deploying digital twins by supporting large-scale pilot programs.
    \item Identify seucurity issues and ensure cybersecurity measures are in place to safeguard the data streams and prevent digital twin manipulation.
\end{itemize}

\subsubsection{Government and Regulatory Bodies}
\textbf{Role}: Provide regulatory frameworks that ensure the safe and secure use of autonomous vessels supported by digital twins. Governments should also offer incentives for research and development in this area.

\textbf{Recommendations}:
\begin{itemize}
    \item Establish clear regulations for autonomous vessels, particularly in regard to collision avoidance and environmental protection.
    \item Support standardization efforts for data-driven modeling, sensor technology integration, and digital twin development to facilitate widespread adoption.
    \item Create funding programs to encourage the maritime industry to adopt green technologies and reduce emissions, leveraging digital twins.
\end{itemize}

\subsubsection{Classification Societies}
\textbf{Role}: Establish safety and performance standards for vessels employing digital twin technologies. These societies will play a crucial role in certifying and ensuring the reliability of digital twins in maritime operations.

\textbf{Recommendations}:
\begin{itemize}
    \item Develop guidelines and standards to certify digital twins in their various capability levels, ensuring safety and performance in all operations.
    \item Implement risk assessment frameworks that incorporate digital twins in predictive maintenance and operational safety assessments.
    \item Promote the adoption of digital twins for real-time condition monitoring to reduce human errors and enhance predictive maintenance.
\end{itemize}

\subsubsection{Research Institutes and Academia}
\textbf{Role}: Conduct fundamental and applied research on digital twin technologies, focusing on areas such as artificial intelligence, advanced control approaches, real-time analytics, sensor and communication technologies, and big data management.

\textbf{Recommendations}:
\begin{itemize}
    \item Explore hybrid models combining physics-based and data-driven approaches to improve real-time decision-making and situational awareness of vessels.
    \item Partner with industry stakeholders to develop new methods for sensor fusion and prediction algorithms, particularly for collision avoidance and path optimization of autonomous vessels.
    \item Research and propose solutions for the ethical and safety concerns arising from autonomous digital twins.
\end{itemize}

\subsubsection{Society in General (Public and Environmental Groups)}
\textbf{Role}: Demand safer, more efficient, and environmentally sustainable maritime operations. Society at large has a stake in ensuring that digital twin technologies reduce emissions and contribute to global environmental goals.

\textbf{Recommendations}:
\begin{itemize}
    \item Advocate for the widespread adoption of digital twin technologies to reduce maritime accidents and environmental impacts.
    \item Engage with maritime technology developers and governments to ensure transparency and accountability in the development of autonomous systems.
    \item Support research on the societal impacts of autonomous maritime systems, ensuring that these technologies create new jobs and safeguard public safety.
\end{itemize}
It is evident from this paper that to fully realize the potential of digital twins in autonomous surface vessels, cross-disciplinary collaboration among all stakeholders is essential. Each group brings unique expertise and resources that contribute to the development and deployment of robust, efficient, and secure digital twin technologies. By fostering close cooperation, stakeholders can address current technological, regulatory, and operational challenges while promoting innovation. This collaborative effort will ensure that digital twins not only enhance the safety and efficiency of maritime operations but also align with broader societal goals of sustainability and environmental responsibility.

\section*{Acknowledgments}
This work is part of SFI AutoShip, an 8-year research-based innovation center. 
In addition, this research project is integrated into the PERSEUS doctoral program. 
We want to thank our partners, including the Research Council of Norway, under project number 309230, and the European Union’s Horizon 2020 research and innovation program under the Marie Skłodowska-Curie grant agreement number 101034240.

\bibliographystyle{elsarticle-harv}
\bibliography{refs}

\end{document}